\documentclass[preprint,12pt]{elsarticle}

\usepackage{amssymb}
\usepackage[utf8]{inputenc}

\usepackage{enumerate}
\usepackage{lineno}
\usepackage{amsmath}
\DeclareMathOperator{\Tr}{tr}
\DeclareMathOperator{\Dev}{dev}
\usepackage{amssymb}
\usepackage{amsfonts}
\usepackage{bm}
\usepackage{miller}
\usepackage{graphicx}
\usepackage{textcomp}
\usepackage{nicefrac}
\usepackage{siunitx}
\usepackage[font=footnotesize,labelfont=bf]{subcaption}
\usepackage{boldline}
\usepackage{overpic}      
\usepackage{stackengine}  %
\usepackage{placeins} 
\usepackage{datetime}
\usepackage{hyperref}
\usepackage{color}
\usepackage{comment}

\usepackage{caption}
\usepackage{subcaption}

\usepackage{geometry}
\journal{}

\begin{document}

\begin{frontmatter}

\title{Simulation-based Super-Resolution EBSD for Measurements of Relative Deformation Gradient Tensors}

\author[AGH,ST]{Aimo Winkelmann}

\affiliation[AGH]{organization={AGH University of Krakow, \mbox{Academic Centre for Materials and Nanotechnology (ACMiN)}},
            addressline={\mbox{al.\@ A. Mickiewicza 30}}, 
            city={\mbox{30-059 Kraków}},
            country={Poland}}

\affiliation[AGH1]{organization={AGH University of Krakow, Faculty of \mbox{Metals Engineering and Industrial Computer Science}},
            addressline={\mbox{al.\@ A. Mickiewicza 30}}, 
            city={\mbox{30-059 Kraków}},
            country={Poland}}

\affiliation[ST]{organization={ST Development GmbH},
            addressline={\mbox{Wilhelmshöhe 7}}, 
            city={\mbox{33102 Paderborn}},
            country={Germany}}
            
\author[AGH]{Grzegorz Cios}
\author[AGH1]{Konrad Perzyński}
\author[AGH]{Tomasz Tokarski}
\author[ST]{Klaus Mehnert}
\author[AGH1]{Łukasz Madej}
\author[AGH]{Piotr Bała}

\begin{abstract}
We summarize a data analysis approach for electron backscatter diffraction (EBSD) which uses high-resolution Kikuchi pattern simulations to measure isochoric relative deformation gradient tensors from experimentally measured Kikuchi patterns of relatively low resolution. 
Simulation-based supersampling of the theoretical test diffraction patterns enables a significant precision improvement of tensor parameters obtained in best-fit determinations of strains and orientations from low-resolution experimental patterns.
As an application, we demonstrate high-resolution orientation and strain analysis for the model case of hardness test indents on a Si(100) wafer, using Kikuchi patterns of variable resolution. 
The approach shows noise levels near $1 \times 10^{-4}$ in the relative deviatoric strain norm and in the relative rotation angles on nominally strain-free regions of the silicon wafer.
The strain and rotation measurements are interpreted by finite element simulations.
While confirming the basic findings of previously published studies, the present approach enables a potential reduction in the necessary pattern data size by about two orders of magnitude. 
We estimate that pattern resolutions in the order of $256\times256$ pixels should be enough to solve a majority of EBSD analysis tasks using pattern matching techniques.
\end{abstract}

\begin{keyword}
Electron Backscatter Diffraction \sep Kikuchi Diffraction  \sep Strain \sep
\end{keyword}
\end{frontmatter}


\section{Introduction}
\label{sec:sample1}

Residual stress and strain are crucial aspects of materials science because they can significantly influence material properties. 
Understanding and controlling both residual stress in polycrystalline materials as well as in strain fields in single crystal thin films is essential for optimizing material properties and predicting component performance in various applications \cite{withers2001mst}.

High-angular resolution EBSD (HR-EBSD) is a powerful tool for analyzing strain in materials due to its high sensitivity to changes in crystal lattice parameters \cite{troost1993apl,wilkinson2006um}.
In addition, by analyzing the local rotational changes which are conveyed by EBSD patterns, HR-EBSD can also be used to study effects of plastic deformation in materials. 

One of the key ideas used in HR-EBSD strain involves comparing the positions of features in experimental EBSD patterns collected from different locations in a sample. 
Small shifts in these features can be directly related to local strain and rotation tensors. 
This method allows for the measurement of elastic strains with a sensitivity of approximately $\pm2 \times 10^{-4}$ \cite{wilkinson2006um} in the components of the infinitesimal strain tensor, making it suitable for studying a variety of phenomena, including residual stresses, strain fields around defects, and for analyzing the behavior of materials under load. 
Similar noise levels of a few $10^{-4}$ \cite{dingley2018icotom,edwards2022sa} have been reported for HR-EBSD measurements on nominally undistorted samples of different materials.

Realistic Kikuchi pattern simulations based on the dynamical theory of electron diffraction \cite{winkelmann2007um} can also be applied to the problem of EBSD from strained crystals. 
The theoretical potential of simulation-based strain tensor extraction from EBSD data has been investigated using numerical studies of simulated EBSD patterns \cite{fullwood2015mc,ruggles2018um,kurniawan2021sm}.
The key role of an independent determination of the projection geometry for accurate determination of strain by EBSD has been illustrated in \cite{kurniawan2021sm}. 
With the use of ideal, simulated data, it has been found that the simultaneous determination of the projection center and deformation tensor severely degrades the precision. 
The observed drop from a basic machine precision $10^{-8}$ under ideal numerical conditions to $10^{-4}$ reflects the fact that the simultaneous fit of PC and deformations (orientation and strain) is ill-conditioned, as has been argued by Alkorta \textit{et al.} \cite{alkorta2013um}.
However, the theoretical investigation in \cite{kurniawan2021sm} suggested that a full pattern matching approach can determine strain tensors based on relatively lower resolution binned patterns. Previously, the influence of the collected EBSD pattern resolution was investigated in \cite{britton2013um2}.
The authors in \cite{zhu2022um} have applied global optimization algorithms to investigate a large parameter space using simulated ideal data with noise.  
They found that deformation state extraction based on simulated patterns showed a mean accuracy of $1 \times 10^{-3}$ in the shear components and about $2 \times 10^{-3}$ in the diagonal components (normal strains) when optimizing while including the projection center in the fit. 
This approach was applied on experimental data in a hybrid algorithm that fits the reference point strain state based on simulated data and then measures the relative strains with a conventional cross-correlation based approach between experimental reference and experimental data patterns. Correcting the relative experimental strains by the simulation based reference strains, partially more consistent strain states were observed in the shear component $\epsilon_{12}$, while the agreement in the other components was less good.

In the present paper we study the application of Kikuchi pattern simulation to the problem of strain determination from real, experimental, EBSD patterns.
Simulation-based supersampling of the theoretical test diffraction patterns enables a significant precision improvement of tensor parameters obtained in best-fit determinations of relative isochoric strains from low-resolution experimental patterns.
For testing new approaches of strain measurement by EBSD, we use hardness tester indents on Si wafers which serve as reproducible reference cases, with published data available in several studies
\cite{wilkinson2012mt,britton2013um2,koko2021arxiv}.
We demonstrate that the approach presented here is able to closely reproduce results obtained previously from full resolution raw patterns (e.g. $1244 \times 1024$ in 16bit resolution) by simulation-based super-sampling of relatively low-resolution 8-bit patterns with $156 \times 128$ or even $78 \times 64$ pixels.
This means that data storage requirements for the same experiment are potentially reduced by up to about two orders of magnitude.  


\section{Principle of Parameter-Super-Resolution EBSD Pattern Matching}

\begin{figure}[htb!]
\begin{center}
    \includegraphics[width=1.0\textwidth]{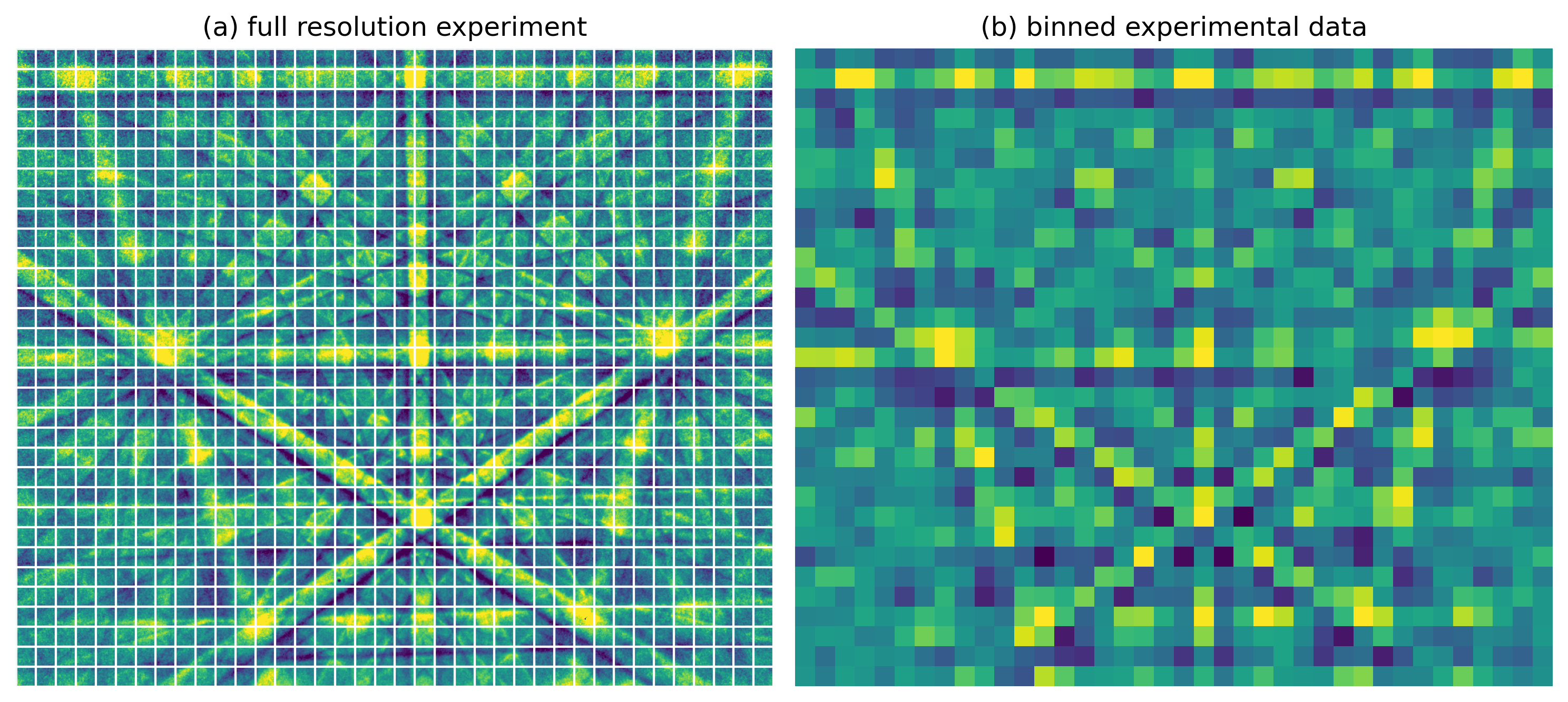}\\
    \includegraphics[width=1.0\textwidth]{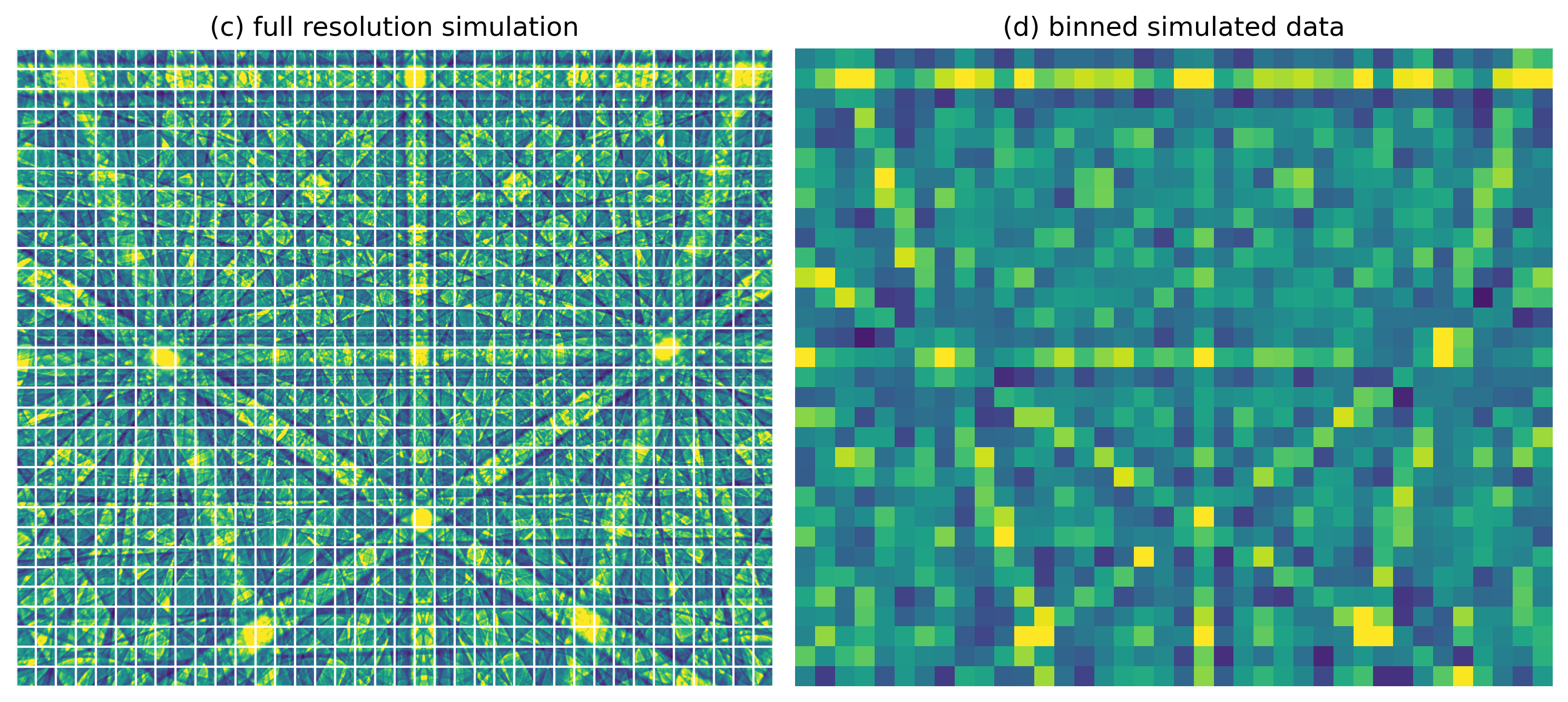}
    \vspace{-3ex}
    \end{center}
        \caption{Principle of super-resolution Kikuchi pattern parameter fitting. Simulated Kikuchi pattern data with $1244\times1024$ pixels on the detector area (c) is used to predict high-resolution parameter changes from $38\times32$ pixel-resolution experimental data (b) via comparison to consistently binned high-resolution simulated data (d), thus circumventing the use of full-resolution $1244\times1024$ experimental data (a).
        In practical use, the tolerable binning will depend on the required resolution of the fitted parameters, which depends on the experimental binning as well as on the number of supersamples in a binned pixel.}
        \label{fig:super}
\end{figure}

The principle of the simulation-based Kikuchi pattern parameter fitting approach can be summarized with reference to Figure \ref{fig:super}. 
Our aim is to work with experimental EBSD patterns optimized for the lowest possible resolution that is compatible with a specific parameter measurement problem (i.e. orientation, strain, crystallographic phase).

As an illustration example, we show a $38\times32$ pixel-resolution experimental measurement in Figure \ref{fig:super}(b), which can be assumed to be produced by an underlying experimental full-resolution intensity distribution shown by $1244\times1024$ pixels on the same detector area in Figure \ref{fig:super}(b). 
We can see that for each of the $38\times32$ detector sectors, the intensity in the $38\times32$ grid cells of the $1244\times1024$ full resolution image is not constant, but is showing internal substructure.
This internal diffraction structure of the large-area pixels in Figure \ref{fig:super}(b) is lost when measuring a low-resolution binned experimental pattern.
A relative comparison of feature shifts (as applied in other HR-EBSD approaches) in two such low-resolution experimental patterns cannot take into account the effects of an underlying, intrinsicallyhigher resolution, experimental distribution.  
However, if we have a theoretical high resolution pattern simulation available as shown in \mbox{Figure \ref{fig:super}(c)}, we can predict precise, fit-parameter dependent, changes of the \textit{mean intensity} in each of the pixel areas of a \textit{binned} high-resolution simulation in Figure \ref{fig:super}(d), which can then be compared to the low-resolution experiment shown in Figure \ref{fig:super}(b).
For this approach, we use the term "\textit{parameter super-resolution}" meaning that the method is designed to infer high-resolution (HR) \textit{parameter} output from low-resolution (LR) \textit{image} input.
Thus the approach can be summarized as being based on the indirect simulation of the combined effects of small shifts of high-resolution intensity features in low-resolution pixels, and not on comparing small shifts of high-resolution pixel intensities directly. 

In practical use, the tolerable lower limit of experimental pattern resolution will depend on the required resolution of the fitted parameters, which not only depends on the experimental binning but also on the number of supersamples in a binned pixel. 
As we will show below using a quantitative error analysis, for some some applications even low-resolution 8-bit patterns with $156 \times 128$ or even $78 \times 64$ pixels can provide similar information as was previously obtained using 16-bit full resolution experimental patterns in the $1000 \times 1000$ pixel range.

\FloatBarrier
\section{Theoretical Background}

\subsection{Projective Transformations}

The parameter fitting approach presented in this study is based on a consistent reprojection of Kikuchi pattern intensities from a simulated, undeformed, master sphere.
The geometrical EBSD pattern analysis pipeline \cite{britton2016mc} describes the gnomonic projection of unit vector directions $[x_M,y_M,z_M]$ from a spherical master Kikuchi diffraction pattern to projected coordinates $[p_1, p_2, p_3]$ on a planar screen.
The relevant mathematics of projective geometry is treated in a number of previous studies applying different strain analysis approaches
\cite{alkorta2017um,winkelmann2018prm,ernould2020amat,ernould2020thesis,ernould2021um,ernould2022mc,ernould2022ch2,shi2021mc,shi2022mc2,shi2024mc}.
A major part of the EBSD pattern geometry can be understood via a projective transformation $\mathbf{F}$ acting on the homogeneous coordinates $[x_M,y_M,z_M]$:

\begin{equation} 
    \begin{bmatrix}
        p_1 \\
        p_2 \\
        p_3  
    \end{bmatrix} = \mathbf{F} \begin{bmatrix}
        x_M \\
        y_M  \\
        z_M  
    \end{bmatrix}
    = \mathbf{R}\,\mathbf{U}\,\begin{bmatrix}
        x_M \\
        y_M  \\
        z_M  
    \end{bmatrix}
    \label{eq:F}
\end{equation}

The coordinates $(x_g, y_g)$ in the standard embedded projection plane at $z_g=1$ are given by dehomogenization of the projective coordinates as $(x_g, y_g) = (p_1/p_3,\, p_2/p_3)$ for all projected points with $p_3 \neq 0$ \cite{winkelmann2020mat}.
The $3 \times 3$ matrix $\mathbf{F}$ has 8 degrees of freedom, because scaling of $\mathbf{F}$ by a real factor does not change the projective transformation \cite{hartley2003MVG}.

\subsection{Strain Tensors}

As has been previously discussed by Maurice \textit{et al.} \cite{maurice2012um,maurice2014rms}, equation (\ref{eq:F}) can also be seen as the definition of the  deformation gradient tensor $\mathbf{F}$ with the polar decomposition $\mathbf{F} = \mathbf{R} \mathbf{U} = \mathbf{V} \mathbf{R}$ \cite{reddy2013}.
Physically, the two decompositions correspond (a) to an initial stretch $\mathbf{U}$ and a subsequent rotation $\mathbf{R}$ of the reference master sphere ($\mathbf{R} \mathbf{U}$), compared to (b) an initial rotation and a subsequent stretch ($\mathbf{V} \mathbf{R})$.
The right Biot stretch tensor $\bm{U}$ \cite{neff2016arma} describes the distortion in the sample system, while the left Biot Stretch Tensor $\bm{V} = \bm{R} \bm{U} \bm{R}^T$ can be used to describe the distortion in the rotated crystallophysical \cite{shuvalov1988mc,haussuehl2007physical} coordinate system.

In the context of EBSD pattern matching approaches, the missing 9th degree of freedom in the theoretical $3\times3$ model matrix $\mathbf{F}$ matches the effect of a severely reduced experimental sensitivity of Kikuchi diffraction patterns to a small uniform expansion of the lattice relative to the changes of angles between crystallographic directions \cite{villert2009jm}.
When the deformation gradient $\mathbf{F}$ acts on the crystal structure, the change in lattice spacings $d_{hkl}$ will lead to a change in Kikuchi band widths $2\theta_{Bragg}$ as a function of $d_{hkl}$. 
However, for the fitting approach we use here, we make the approximation that in the relevant EBSD regime of small strains, we are \textit{not} sensitive to the related minimal changes in Kikuchi band widths, and thus we can only determine changes in shape (distortions) of a reference crystal but not changes in size (dilations).
This is why we constrain $\mathbf{U}$ to be an isochoric transformation, i.e. we prescribe a unit determinant for all the matrices involved $\det{\mathbf{F}}=\det{\mathbf{R}} = \det{\mathbf{U}}=1$, thereby having 8 free parameters in the $3\times3$ matrix $\mathbf{F}$.
The ideal, simplified model of equation \ref{eq:F}, can be straightforwardly extended to include additional experimental boundary conditions in the projection pipeline, such as sample-to detector rotation or optical disortions \cite{shi2022mc2}.

\subsection{Finite and Infinitesimal Strain}

The Biot stretch tensor $\bm{U}$ and the corresponding Biot strain tensor $\bm{E}_B = \bm{U} - \bm{I}$ \cite{biot1939pm} are central to the pattern-matching based strain fitting approach discussed in the present paper.
In order to be able to make connections to infinitesimal strain frameworks which are in common use for the interpretation of EBSD data \cite{wilkinson2006um,jackson2016mm,ernould2020amat}, we will discuss the relationship between the infinitesimal strain tensor $\bm{\varepsilon}$ and the finite Biot stretch tensor $\bm{U}$.

In the material (Lagrangian) description of deformation in continuum mechanics \cite{reddy2013}, the infinitesimal strain tensor $\bm{\varepsilon}$ can be derived by a linearization of the \mbox{Green-Lagrange} finite strain tensor $\bm{E}$, written in terms of the right \mbox{Cauchy–Green} deformation tensor $\bm{C}$, which is the square of the right Biot stretch tensor $\bm{U}$ (using that $\bm{U}$ is symmetric $\bm{U}^T = \bm{U}$, and that $\bm{R}$ is orthogonal $\bm{R}^T = \bm{R}^{-1}$ in the polar decomposition $\bm{F} = \bm{R}\, \bm{U}$):

\begin{equation}
\bm{C} =  \bm{F}^T\bm{F} = \bm{U}^T \bm{R}^T \bm{R} \bm{U} = \bm{U}^2
\end{equation}

In order to see the role of the Biot strain tensor $\bm{E}_B = \bm{U} - \bm{I}$ \cite{neff2016arma}, we write down the \mbox{Green-Lagrange} finite strain tensor $\bm{E}$ which is defined as:

\begin{equation}
    \bm{E} = \frac{1}{2} (\bm{C}-\bm{I}) =  \frac{1}{2} (\bm{U}^2-\bm{I}) =  \frac{1}{2}((\bm{E}_B + \bm{I})^2 -\bm{I}) 
\end{equation}

Expanding the square leads to:

\begin{equation}
    \bm{E} =  \frac{1}{2} (\bm{E}_B^2 + 2 \bm{E}_B + \bm{I}^2 - \bm{I})
\end{equation}

For small deformations, we approximate $\bm{E}$ using the terms linear in $\bm{E}_B$, giving the infinitesimal strain tensor $\bm{\epsilon}$:  

\begin{equation}
    \epsilon \approx \bm{E}_B = \bm{U} - \bm{I}
    \label{eq:infstrain}
\end{equation}

This shows how the small-strain infinitesimal strain tensor $\epsilon$ can be derived either from the finite \mbox{Cauchy–Green} deformation tensor $\bm{C}$ or directly from the finite Biot strain tensor $\bm{E}_B$. If data interpretation in an infinitesimal framework is relevant, the infinitesimal strain tensor matrix elements $\epsilon_{ij}$ can be directly approximated by the Biot strain tensor elements $\epsilon_{ij} \approx [\bm{E}_B]_{ij}$.

In many continuum mechanics problems, the Biot strain tensor $\bm{E}_B$ is not directly experimentally available, but has to be calculated from the finite \mbox{Cauchy–Green} deformation tensor $\bm{C}$.
In the forward simulation approach implemented in this study, however, the deformation gradient tensor $\bm{F}$ is synthesized from $\bm{R}$ and $\bm{U}$, and the matrix elements $[\bm{E}_B]_{ij}$ are directly entering as fit parameters.
In the transformation of pixel intensities from the undeformed to the deformed configuration, the deformation gradient tensor $\bm{F} = \bm{R}\,\bm{U} = \bm{R}\,(\bm{E}_B + \bm{I}) \approx \bm{R}\,(\epsilon + \bm{I})$ is acting in its second role as a projective transformation matrix of crystallographic directions, which is the basis of the extraction of strain parameters from a fit of simulated distorted data to experimental diffraction patterns. 
The experimental challenge is that $\bm{F} = \bm{R}\,\bm{U}$ as it is written is a severe compression of the complete projection pipeline into only 9 effective parameters, in which the strain parameters are entering together with other geometrical parameters.
A key problem of experimental strain determination by EBSD is to fix these additional geometrical parameters well enough, so that the local strain can be extracted correctly. 

\subsection{Characterization of Shape Changes}

The strain tensor elements fitted to EBSD Kikuchi patterns can provide direct physical insight if the coordinate system for the best representation of the tensor can be chosen based on prior knowledge about the deformation \cite{edwards2024nt}.
If this is not the case, tensor invariants provide important information which is independent of the chosen coordinate system. 
For example, the principal strains and their directions can be helpful in crystallographic interpretations of a deformation \cite{dingley2018icotom}.
In the present study, we use a set of invariants of the Hencky strain tensor $\bm{E}_H  = \ln{\bm{U}}$ \cite{neff2016arma} to characterize the changes in size and shape of a material element which are caused by a strain tensor independently of the orientation of the cartesian axes. 

As described in \cite{criscione2000jmps}, a change in size (dilation) can be characterized by an invariant $K_{1}$: 
\begin{equation}
    K_{1} = \Tr \bm{E}_H = \ln{J} \approx \Tr \bm{E}_B
    \label{eq:K1}
\end{equation}
where $J = \det{\bm{F}} = \det{\bm{U}}$ is the volume ratio of a material element after the deformation relative to the reference configuration.
For the isochoric deformations considered in the fitting approach of this study, we have $J = 1$ and thus a trivial constant $K_{1} = 0$ which gives no further information.

In comparison to $K_1$, which simply describes a constant unit cell volume in the present case, quantitative measures of distortion (= isochoric deformation) of that unit cell can be constructed using a normalized deviator $\bm{\Phi}$: 
\begin{equation}
    \bm{\Phi} =  \Dev{\bm{E}_H}\, / K_2
\end{equation}

where the invariant $K_2$ is defined using the Frobenius norm $||\ldots||_2$, and the corresponding approximations for small strains:

\begin{equation}
    K_{2} = || \bm{E}_H||_2 \approx || \bm{E}_B||_2 \approx || \bm{\varepsilon}||_2 
    \label{eqn:K2}
\end{equation}

In this way, the Hencky strain tensor can be decomposed into $K_2\geq 0$ which describes how much the shape of the unit cell changes, and $\mathbf{\Phi}$, which specifies how the shape change looks like:
\begin{equation}
    \bm{E}_H = K_2\,\bm{\Phi}
\end{equation}

\begin{figure}[htb!]
\begin{center}
    \includegraphics[width=.75\textwidth]{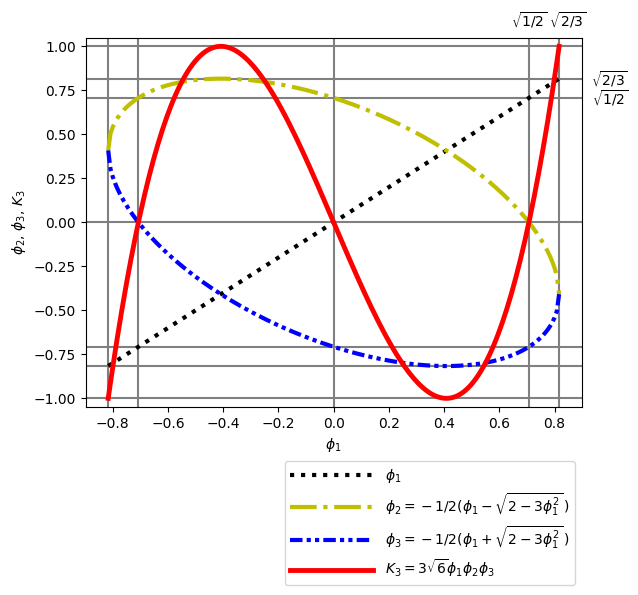}
    \vspace{-3ex}
    \end{center}
        \caption{Principal strains $\phi_1$, $\phi_2$, $\phi_3$ of the normalized deviator $\bm{\Phi}$, and the resulting invariant $K_3$ (eqn. \ref{eqn:K3}) which characterizes the type of distortion. $\phi_1$, $\phi_2$, $\phi_3$ are constrained by the conditions $\phi_1 + \phi_2 + \phi_3 = 0$ (trace of $\bm{\Phi}$) and $\phi_1^2 + \phi_2^2 + \phi_3^2 = 1.0$ (norm of $\bm{\Phi}$), leading to the expressions for $\phi_2$ and $\phi_3$ and the determinant $\det\bm{\Phi} = \phi_1 \cdot \phi_2 \cdot \phi_3$ as a function of  $\phi_1$.}
        \label{fig:k3}
\end{figure}

Moreover, the type of distortion can be numerically characterized by the scaled determinant $K_{3}$ of the normalized deviator $\mathbf{\Phi}$:
\begin{equation}
    K_{3} = 3\sqrt{6}\det \bm{\Phi}
    \label{eqn:K3}
\end{equation}
The invariant $K_{3}$ is in the range $[-1,1]$, with the extreme cases $K_3=1$ for uniaxial extension (= equibiaxial contraction), $K_3=-1$ for uniaxial contraction (= equibiaxial extension), and $K_3=0$ for pure shear strain, where one principal axis is not stretched at all and the other two principal axes are, respectively, stretched and compressed in a reciprocal way.
In terms of the eigenvalues $\lambda_1, \lambda_2, \lambda_3$ of the stretch tensor $\bm{U}$, pure shear strain is a special case of an isochoric deformation,
with the principal strains related by a condition like $\lambda_1 = \lambda,\,\, \lambda_2 = 1/\lambda,\,\, \lambda_3 = 1$   \cite{thiel2019ijnlm}.
We note that the mode of distortion $K_3$ becomes irrelevant when there is actually no distortion (i.e. $K_2 \approx 0$).
Moreover, in the presence of noise levels which are in the same order as the strain norm itself, the type of distortion inferred from the noisy strain tensor elements will also become random and the value of $K_3$ will then randomly vary from map point to map point.
This is qualitatively similar to the way the direction of local rotation axes would cause orientation IPF-color noise for small the rotation angle near the noise level \cite{prior1999jm}.
We note that a certain precision stated for the strain tensor norm implies a significantly higher limit on how sure we are about the type of distortion that is present at a certain absolute magnitude of distortion.
Our measurements below indicate that for strain norm noise of 0.1 mm/m, the type of distortion already becomes undetermined for strain norms of about 1 mm/m and smaller.

\subsection{Relative Deformation Gradients}

As discussed in \cite{winkelmann2021jm}, the raw fit of the deformation gradient tensor $\bm{F}$ through consistently deformed simulated Kikuchi patterns will show a significant bias due to the missing treatment of excess-deficiency effects in experimentally measured Kikuchi patterns by the pattern simulation model, among other influences.
However, we make the assumption that the \textit{ relative} deformation gradients, i.e. relative to specified reference points, can still be well approximated via the relative changes obtained via the (biased) fits to simulated, deformed Kikuchi patterns.
In order to estimate under which conditions this assumption holds, we model the experimentally fitted deformation gradient tensor $\bm{F}^{E}$ as the result of an unknown deformation bias $\bm{F}^{B}$ that acts before the true deformation gradient $\bm{F}$:

\begin{equation}
    \bm{F}^{E} = \bm{F} \bm{F}^{B}
\end{equation}

Which gives for the true deformation gradient (with $\bm{F}^{B}$ unknown):
\begin{equation}
    \bm{F} = \bm{F}^{E} [\bm{F}^{B}]^{-1}
\end{equation}

Furthermore, the relative deformation gradient $\bm{F}_{10} $ at a point 1 relative to a reference point 0 can be defined as:

\begin{equation}
    \bm{F}_{1} = \bm{F}_{10} \bm{F}_{0}
\end{equation}

For the relative, experimental deformation gradient tensor $\bm{F}_{10}^{E}$ between the fitted deformation gradients $ \bm{F}^{E}$ at points 0 and 1 this leads to:

\begin{equation}
   \bm{F}_{10}^{E} = \bm{F}_{1}^{E} (\bm{F}_{0}^{E})^{-1}
\end{equation}

We separate the experimental values into the biases and the true values:

\begin{equation}
    \bm{F}_{10}^{E} = \bm{F}_{1} \bm{F}^{B}_{1}  [\bm{F}^{B}_{0}]^{-1}  [\bm{F}_{0}]^{-1} 
\end{equation}

If we can assume that the bias deformation gradient is approximately constant for the two observation points, $\bm{F}^{B}_{0}  \approx \bm{F}^{B}_{1}  = \bm{F}^{B}$ we have the following:

\begin{equation}
    \bm{F}_{10}^{E} \approx \bm{F}_{1} [\bm{F}_{0}]^{-1}  = \bm{F}_{10}
\end{equation}

showing that we can still obtain an estimation of the true relative deformation gradient from the observed, biased, deformation gradients if the bias deformation is approximately constant at the two observation points.

\FloatBarrier
\section{Experimental Details}

\subsection{EBSD Measurements and Data Analysis}

The EBSD setup applied in this study consists of a field emission gun SEM Versa 3D (FEI), which was equipped with a lensless, fiber-optic-based Symmetry S2 EBSD detector (Oxford Instruments Nanoanalysis). 
The SEM beam voltage was 20kV, the beam current was approx. 50 nA, and the sample tilt was $70^{\circ}$.
The EBSD detector was operated in "Resolution" mode acquiring patterns with $622 \times 512$ and $1244 \times 1024$ pixel resolution using the Aztec 6.1 SP2 acquisition software (Oxford Instruments Nanoanalysis). 
The acquired data was exported to the hdf5-based, open data format H5OINA \cite{h5oina2024} for further analysis.

The indents were made using a Wilson Tukon 1202 hardness tester using a Vickers diamond tip with a load of 50 gf (gram-force, ISO 6507) on a 10x10x1 mm$^3$ (001) silicon single crystal. These conditions are similar to previous studies on indenter strains that have been published in the literature \cite{wilkinson2012mt,britton2013um2,koko2021arxiv}.

\begin{sloppypar}
Data analysis was carried out with dynamic template matching \cite{trimby2024um} implemented in \mbox{AztecCrystal} \mbox{MapSweeper} (Oxford Instruments Nanoanalysis) using the pattern-matching strain refinement mode available in version 3.3 of \mbox{AztecCrystal}.
The projection center parameters (PCX, PCY, DD) for the maps the were calibrated by a fit of an affine projection of a regular scan grid on a planar surface to 16 independently fitted (for orientation and PC) reference points along the edges of the maps, far away from the indents in nominally strain-free regions \cite{winkelmann2020mat}.  
Additional processing of the H5OINA data for error analysis was performed using custom Python scripts. 
\end{sloppypar}

\subsection{FEM Simulations}

\begin{figure}[htb!]
\begin{center}
    \includegraphics[width=.5\textwidth]{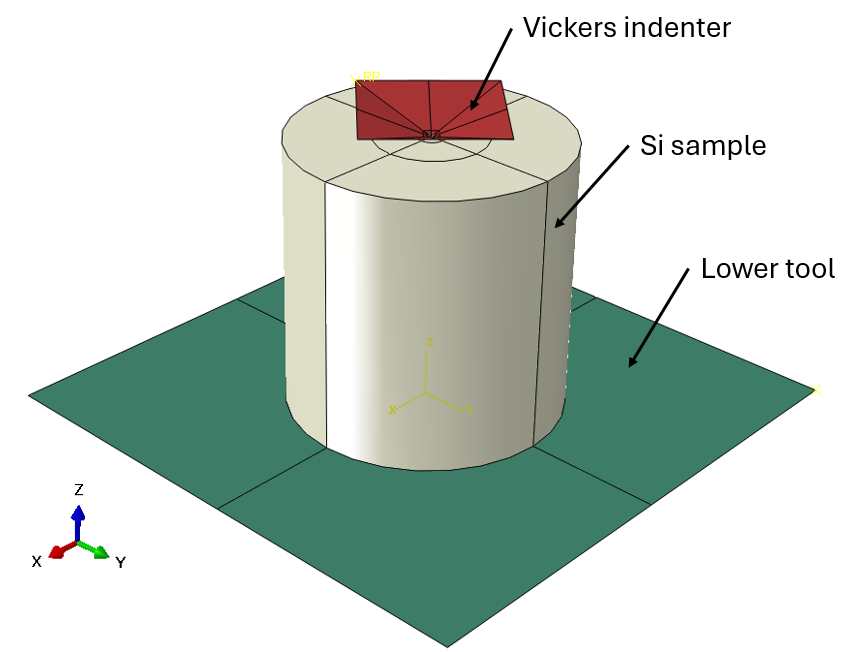}
    \vspace{-3ex}
    \end{center}
        \caption{Assembly of the 3D nanoindentation test.}
        \label{fig:FEM_assembly}
\end{figure}

A numerical model of the nanoindentation test was developed within the commercial Abaqus software. The size of the sample was selected based on \cite{PERZYNSKI201934} as 100 $\mu$m in height and 100$\mu$m radius to avoid any unphysical behaviour. The sample bottom surface was fixed in all directions to reflect the stiff lower tool. The Vickers-type indenter was assumed to be fully rigid during the simulation. The developed model assembly is shown in Figure \ref{fig:FEM_assembly}.

\begin{figure}[htb!]
\begin{center}
    \includegraphics[width=.5\textwidth]{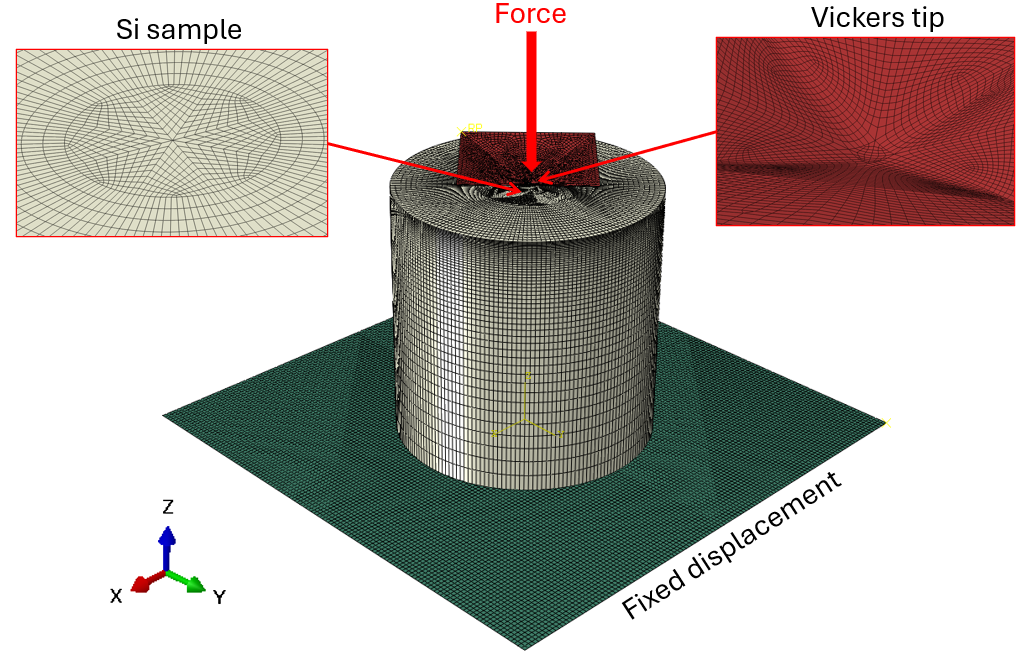}
    \vspace{-3ex}
    \end{center}
        \caption{Finite elements discretization of the nanoindentation test with applied boundary conditions.}
        \label{fig:FEM_mesh}
\end{figure}

The solid 8-node elements with a single Gaussian integration point (C3D8R) and hourglass control were used during the analysis. The number of elements for the sample discretization was set to 160,000. The Vickers indenter and lower tool were discretized using 24,000 and 15,000 4 node rigid shell elements, respectively. The number of finite elements in the model was selected by an extensive mesh sensitivity analysis prior to the simulations. The FE mesh was additionally refined in the area deformed by the Vickers indenter tip, as seen in Figure \ref{fig:FEM_mesh}.

The classical J2 plasticity with Swift law-type hardening curve was used to describe the Si sample behaviour under loading conditions. The hardening model parameters were acquired from \cite{PERZYNSKI20172191}.

\FloatBarrier
\clearpage
\section{Results \& Discussion}

\subsection{Strain and Rotation Tensor Mapping near Indents on Si(100)}

\begin{figure}[t!]
\begin{center}
    \includegraphics[width=.995\textwidth]{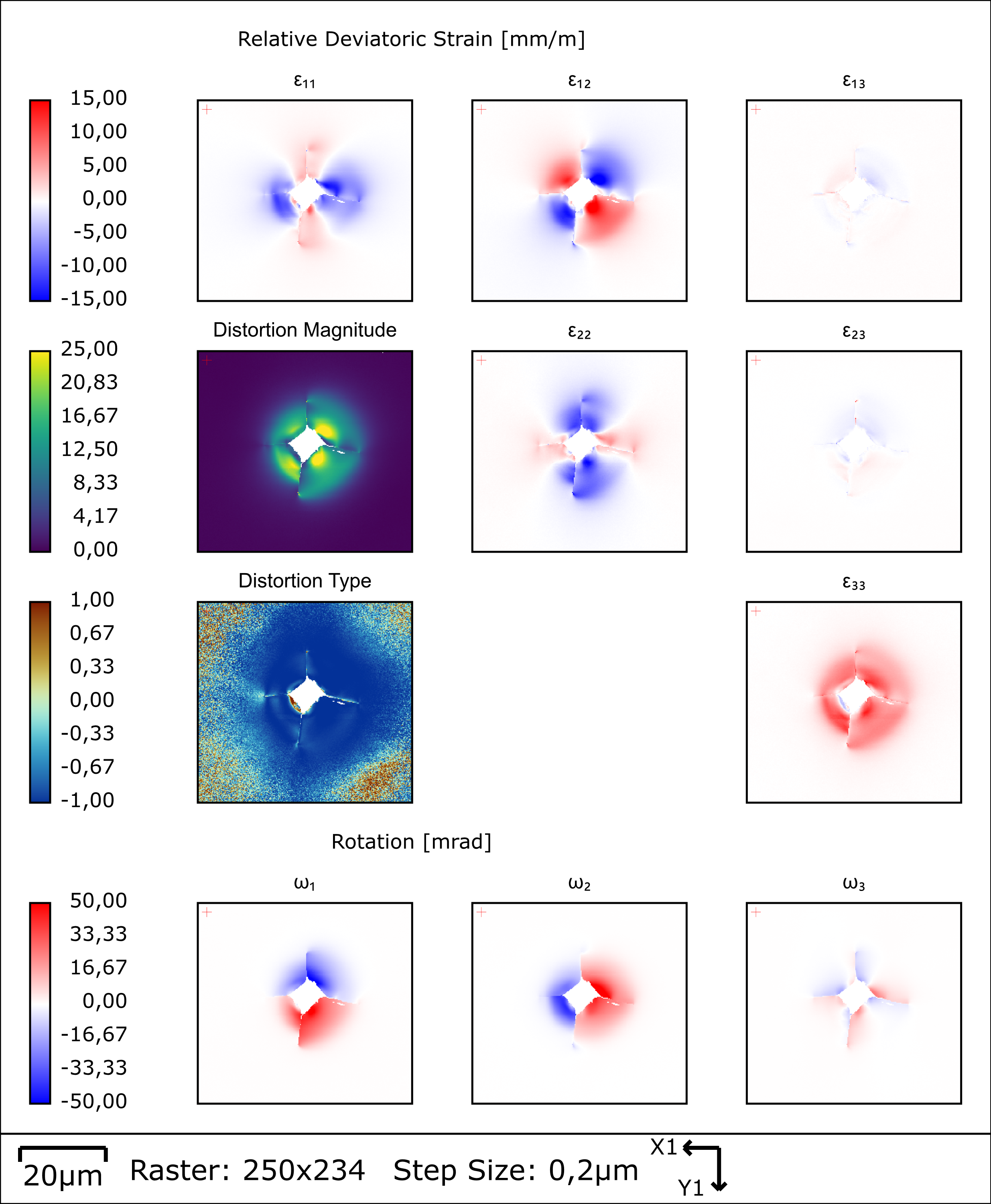}
    \vspace{-3ex}
    \end{center}
        \caption{Reference result with a pattern resolution of $622\times512$ detector pixels, and $3\times3$ simulation supersampling.
        Analysis of relative deviatoric strain for an Si(100) indent. EBSD pattern resolution , Distortion Magnitude = $K_2$ (equation \ref{eqn:K2}) in mm/m, Distortion Type = $K_3$ (equation \ref{eqn:K3}) (dimensionless)}
        \label{fig:si512}
\end{figure}

In Figure \ref{fig:si512}, we show the reference, high-resolution, result for a scan of $250\times234$ map points near an indent on a Si(100) wafer, and the experimental conditions are similar as used in \cite{wilkinson2012mt,britton2013um2}.
The resolution of the base EBSD pattern was $622\times512$ pixels, with $3\times3$ simulation supersampling. 
The fit of the strain and orientation tensors was considered converged when the normalized cross-correlation coefficient $R$ between simulation and experiment changed by less than a convergence threshold of $\Delta R_C < 1\times10^{-6}$. 
We used a PC with a AMD Ryzen Threadripper PRO 5965WX CPU and two NVIDIA RTX4090 GPUs, which gives speeds near 20 patterns per second for the strain refinement, leading to a total analysis time in the order of half an hour for the complete EBSD scan.
Due to the comparable total number of numerical operations, the time required for data analysis remains in a similar order of magnitude for all scenarios of combined binning and supersampling.

For visualization of strain tensor data, we applied color scales that aim to be visually straightforward, i.e., they should easily allow identifying positive, negative, and near-zero results \cite{crameri2020ncomm,crameri2021}. 
The jet color scale can be challenging to interpret when used to plot EBSD strain tensor maps with positive and negative signals \cite{wilkinson2012mt,britton2013um2,vaudin2015um,dingley2018icotom}.

In Figure \ref{fig:si512} we find excellent agreement with previously published data \cite{wilkinson2012mt,britton2013um2,koko2021arxiv} both in the qualitative distributions seen in the strain tensor element maps and in the quantitative magnitude of the signals in the measured strain fields.

\begin{figure}[t!]
\begin{center}
    \includegraphics[width=.995\textwidth]{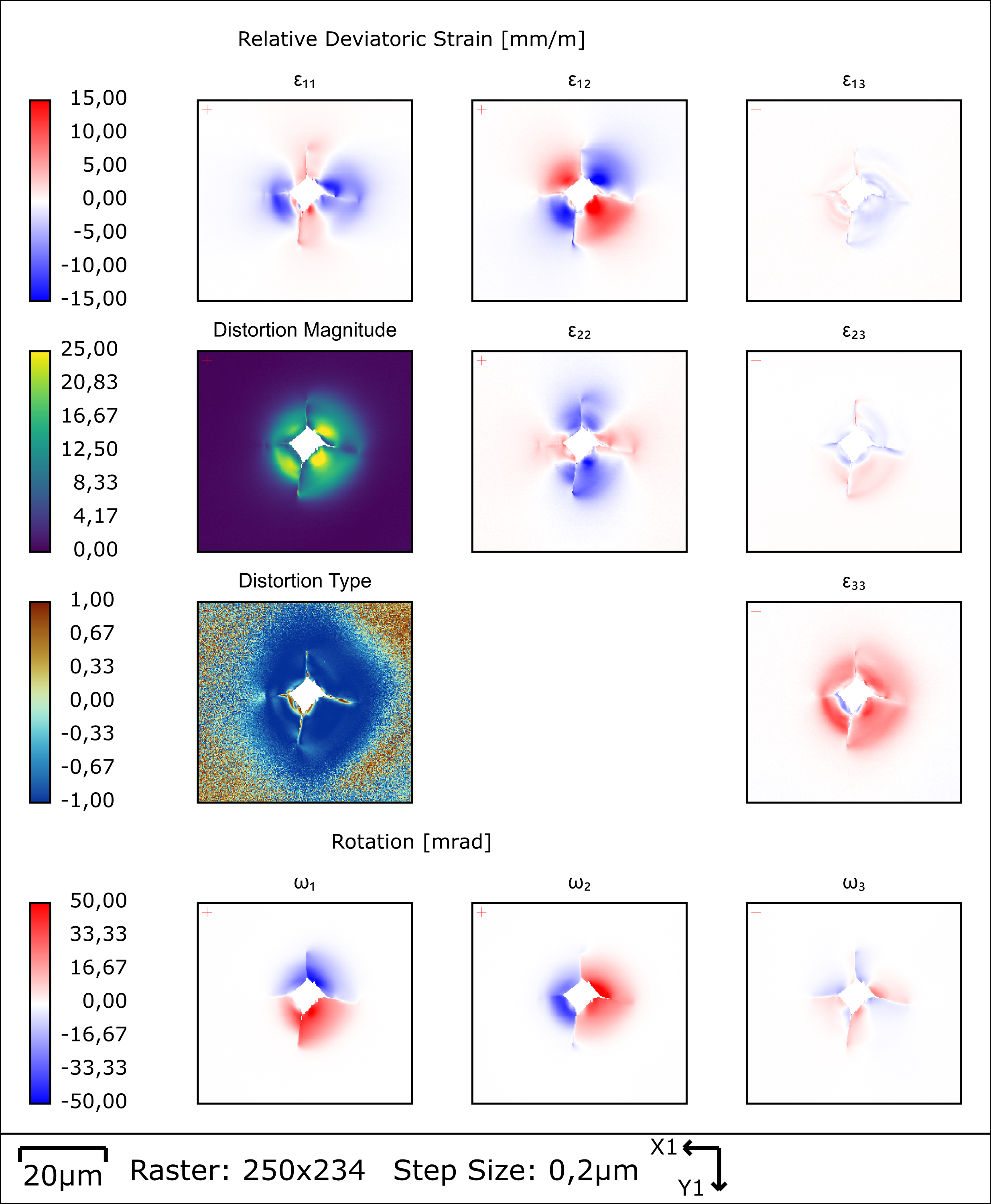}
    \vspace{-3ex}
    \end{center}
        \caption{Relative deviatoric strain for an Si(100) indent, EBSD pattern resolution $155\times128$ (binned), with 
        $13\times13$ supersampling. The other parameters are the same as in Figure \ref{fig:si512}.}
        \label{fig:si128}
\end{figure}

Demonstrating the potential of supersampling a lower resolution experimental pattern with a high resolution simulation, in Figure \ref{fig:si128} we analyzed the same experimental measurement as above, but with the EBSD patterns binned to a resolution of $155\times128$, while using a correspondingly higher $13\times13$ simulation supersampling.

Comparing both results, we find that they are visually nearly identical. The $155\times128$ data shows the largest differences in the maps of $\varepsilon_{13}$ and $\varepsilon_{23}$, which are much smaller than the other components. 
This is consistent with the expectation that the $155\times128$ low pattern resolution data should be sufficient to analyze relatively large strains, while the limitations of this pattern resolution will be seen when analyzing lower magnitude signals.
We quantify these assumptions in the following section by a numerical error analysis of the results obtained at different pattern resolutions.

\FloatBarrier
\clearpage

\FloatBarrier
\clearpage

\FloatBarrier
\clearpage
\subsection{Estimation of Precision and Accuracy}

We can estimate the noise in the relative orientation and strain tensor elements from neighbor-pair differences, if the deformation can be assumed to be smooth, i.e. almost constant, between pairs of neighboring data points. 
For the difference of two Gaussian random variables $\epsilon$, the variances add and an underlying constant mean value is removed via
$\Delta\epsilon_m = \epsilon_{2m+1} - \epsilon_{2m}$, with $m$ indexing the successive unique pairs of independent data points \cite{shumway2017}.

The estimated noise-related standard deviation $\sigma_{\epsilon}$ of the values $\epsilon$ at a single measurement point results as:

\begin{equation}
    \sigma_{\epsilon} = \frac{\sigma_{\Delta\epsilon}}{\sqrt{2}}
\end{equation}

and similar for the rotation angle $\vartheta$:
\begin{equation}
    \sigma_{\vartheta} = \frac{\sigma_{\Delta\vartheta}}{\sqrt{2}}
\end{equation}

\begin{figure}[htb!]
\begin{center}
    \includegraphics[width=.995\textwidth]{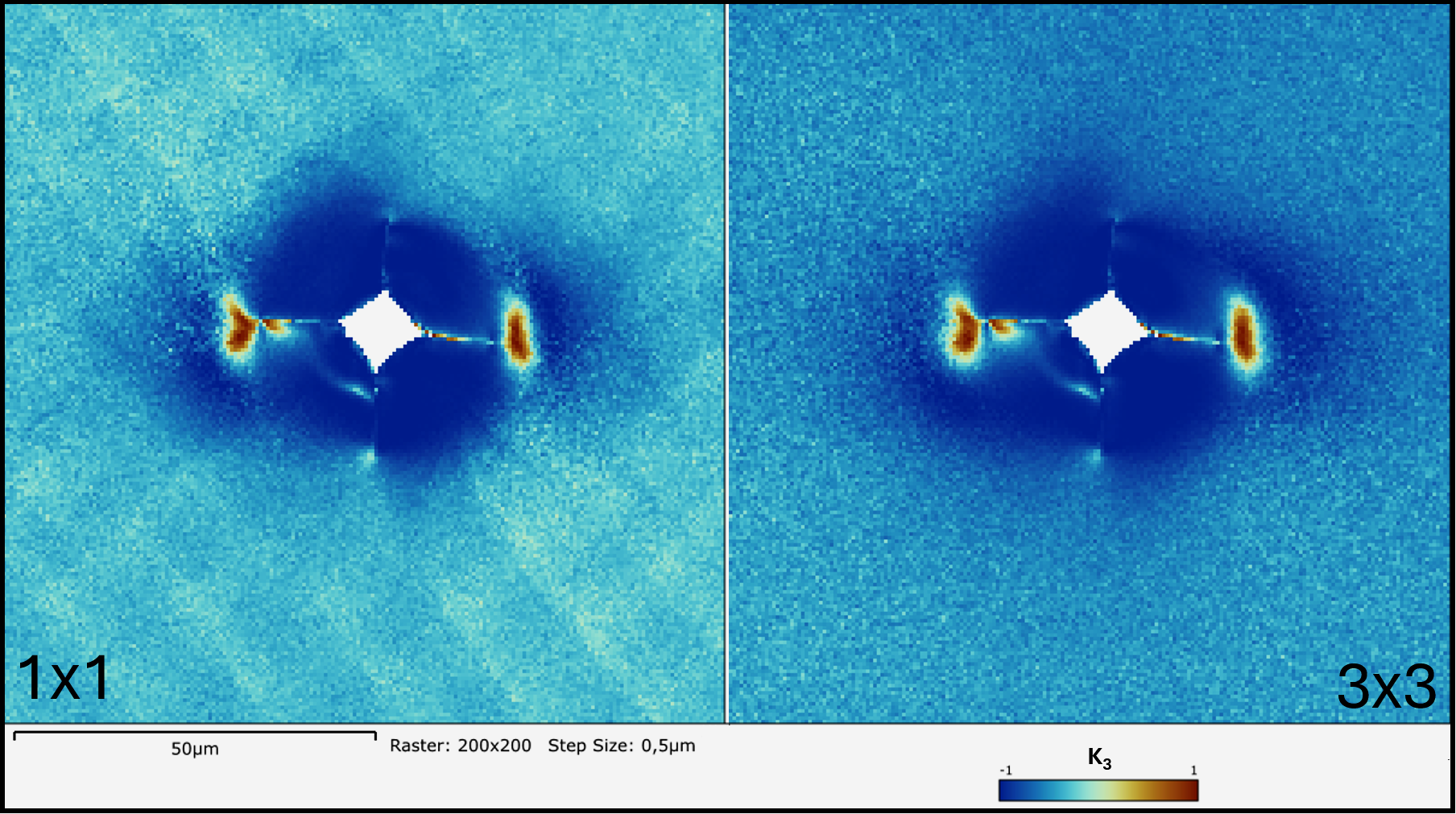}
    \vspace{-3ex}
    \end{center}
        \caption{$K_3$ Distortion type maps (equation \ref{eqn:K3}) plotted from the fit result for the non-relative isochoric strain, pattern resolution $1244\times1024$. The result without supersampling ($1\times1$, left) shows artefacts and while using $3\times3$ (right) simulation supersampling these artifacts are absent. 
        The overall blue color, indicating uniaxial compression seen even in the nominally unstrained area shows the effect of the bias that is introduced by the simulation-model (i.e. neglecting excess-deficiency effects).}
        \label{fig:1024ss}
\end{figure}

In the comparison shown in Table \ref{tab:resu}, we estimate the noise and relative absolute errors of the strain norm and of the rotation angles for different pattern resolutions and supersampling settings, using a measurement with a base resolution of $1244 \times 1024$ pattern pixels and $200\times200$ map points around the Si indent. 
The noise standard deviations $\sigma_{\epsilon}$ (mm/m) and $\sigma_{\vartheta}$ (mrad) were determined from successive pairs of map points along row 10 of the map (i.e. giving 100 difference values), well outside the indent area in the map, which is shown in Figure \ref{fig:1024ss}.
In the statistical analysis of the absolute RMSE differences, we included all signal data points characterized by absolute strain norms $||\epsilon||>1.0$ mm/m and $||\vartheta||>1.0$ mrad, respectively.
In all cases, the indexing threshold was $R_{\mathrm{min}}=0.4$, the convergence threshold $\Delta R_C = 10^{-6}$

In Table \ref{tab:resu}, the root mean-square differences of the strain norm $\epsilon_{\mathrm{RMSE}}$ and of the rotation vector norm (=rotation angle)  $\vartheta_{\mathrm{RMSE}}$ are relative to the reference result for $1244 \times 1024$ pixels and $3 \times 3$ supersampling.
To explain the use of $3\times3$ simulation supersampling even on the $1244 \times 1024$ full resolution pattern, we show the reduction of artifacts in the very low strain regions away from the indents when using a $3\times3$ simulation supersampling, as seen in Figure \ref{fig:1024ss}.
The $K_3$ Distortion type maps (equation \ref{eqn:K3}) plotted from the raw fit result for the non-relative isochoric strain without supersampling on the left side of Figure \ref{fig:1024ss} shows regular artifacts which are probably due to the discrete pixel resolution used in both experiment and simulation. 
When using $3\times3$ simulation supersampling, as shown in the right part of Figure \ref{fig:1024ss}, these artifacts are largely absent. 
However, the overall blue color, which seems to indicate a uniaxial compression even in the nominally unstrained area shows the effect of the bias that is introduced by the simulation-model, with the approximation of neglecting excess-deficiency effects.
This bias is one of the reasons why the raw, non-relative, fit result is of limited value. 
As was discussed above, we thus restrict the analysis of the strain data to the relative strains which we can still obtain even in the presence of bias, if the bias is approximately constant.

The results in Table \ref{tab:resu} indicate that a resolution of $311\times256$ with $7 \times 7$ supersampling still shows about the same levels of noise $\sigma_{\epsilon}$ and $\sigma_{\vartheta}$ as the $1244\times1024$ reference result, while the the $77\times64$ resolution leads to values that are already about twice the reference noise value.

At the same time, the absolute errors $\epsilon_{\mathrm{RMSE}}$ and $\vartheta_{\mathrm{RMSE}}$ at $311 \times 256$ with $7 \times 7$ supersampling relative to the reference result are still in the order of $2 \times 10^{-4}$, while the the $77\times64$ resolution shows an increase of the RMS error by a factor 4 to 5.

As would be expected, when the pattern resolution becomes too low, the loss of spatial information cannot be compensated for anymore by a higher supersampling, and the $77 \times 64$ and $38 \times 32$ resolution show significantly higher noise and absolute errors.
However, depending on the application, even the limited precision delivered by the $38 \times 32$ resolution could be sufficient, for example, for indexing and grain size analysis.

\begin{table}[htb!]
    \centering
    \begin{tabular}{cccccc}
         \hline
         pattern resolution& supersampling &  $\sigma_\epsilon$ (mm/m) & $\sigma_\vartheta$ (mrad) & $\epsilon_{\mathrm{RMSE}}$ (mm/m)   & $\vartheta_{\mathrm{RMSE}}$ (mrad)\\
         \hline
         $1244 \times 1024$ & $3 \times 3$ & 0.073 & 0.046 &  0 & 0 \\
         \hline
         $1244 \times 1024$ & $1 \times 1$ & 0.088 & 0.044 & 0.132 & 0.078\\
         $622 \times 512$& $3 \times 3$    & 0.079 & 0.043 & 0.090 & 0.065\\
         $311 \times 256$& $7 \times 7$    & 0.071 & 0.048 & 0.216 & 0.164\\
         $155\times 128$& $13 \times 13$   & 0.096 & 0.052 & 0.616 & 0.492\\
         $77 \times 64$& $25 \times 25$    & 0.139 & 0.089 & 0.865 & 0.937\\
         $38 \times 32$& $49 \times 49$    & 0.249 & 0.181 & 1.831 & 3.290\\
         \hline
    \end{tabular}
    \caption{
    Estimation of noise and relative absolute errors of strain norm and rotation angles for different pattern resolutions and supersampling settings.
    Noise standard deviations $\sigma_{\epsilon}$ (mm/m) and $\sigma_{\vartheta}$ (mrad), determined from a nominally strain free line 10, well outside the indent area in the map.   
    Root mean-square differences RSME of the strain norm $\epsilon_{\mathrm{RMSE}}$ and of the rotation vector norm (=rotation angle)  $\vartheta_{\mathrm{RMSE}}$ are relative to the reference result for $1244 \times 1024$ pixels and $3 \times 3$ supersampling, for signal data points characterized by absolute strain norms $||\epsilon||>1.0$ mm/m and $||\vartheta||>1.0$ mrad, respectively.
    In all cases, the indexing threshold was $R_{\mathrm{min}}=0.4$, the convergence threshold $\Delta R_C = 10^{-6}$.
    }
    \label{tab:resu}
\end{table}

\begin{figure}[htb!]
\begin{center}
     \centering
     \begin{subfigure}[b]{0.7\textwidth}
        \centering
        \includegraphics[width=\textwidth]{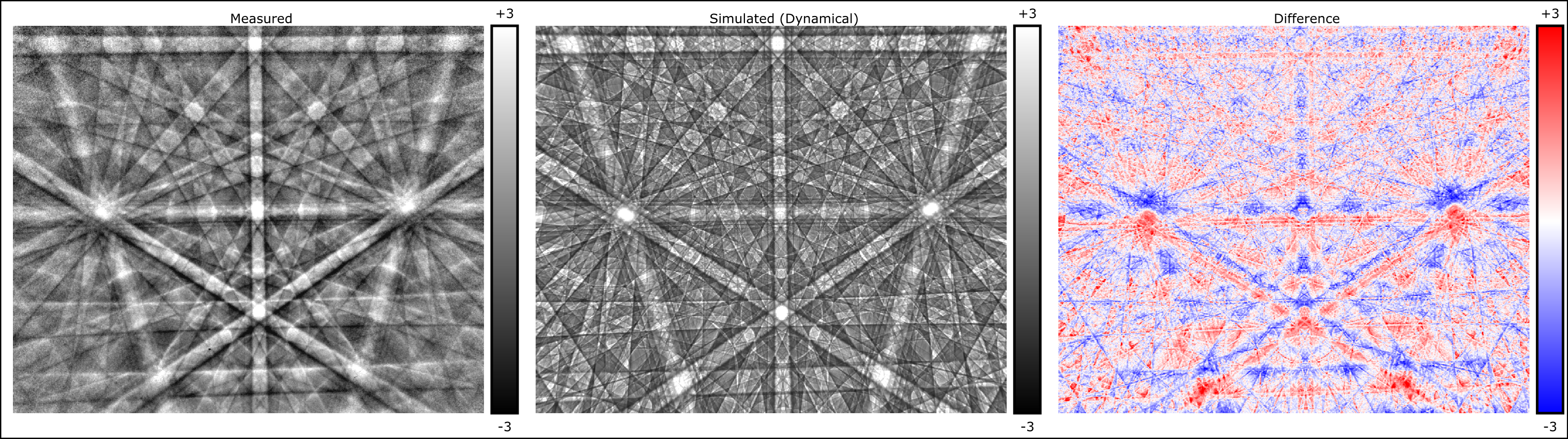}
        \caption{1244x1024 3x3 R=0.6262}
     \end{subfigure}
     \begin{subfigure}[b]{0.7\textwidth}
         \centering
    \includegraphics[width=\textwidth]{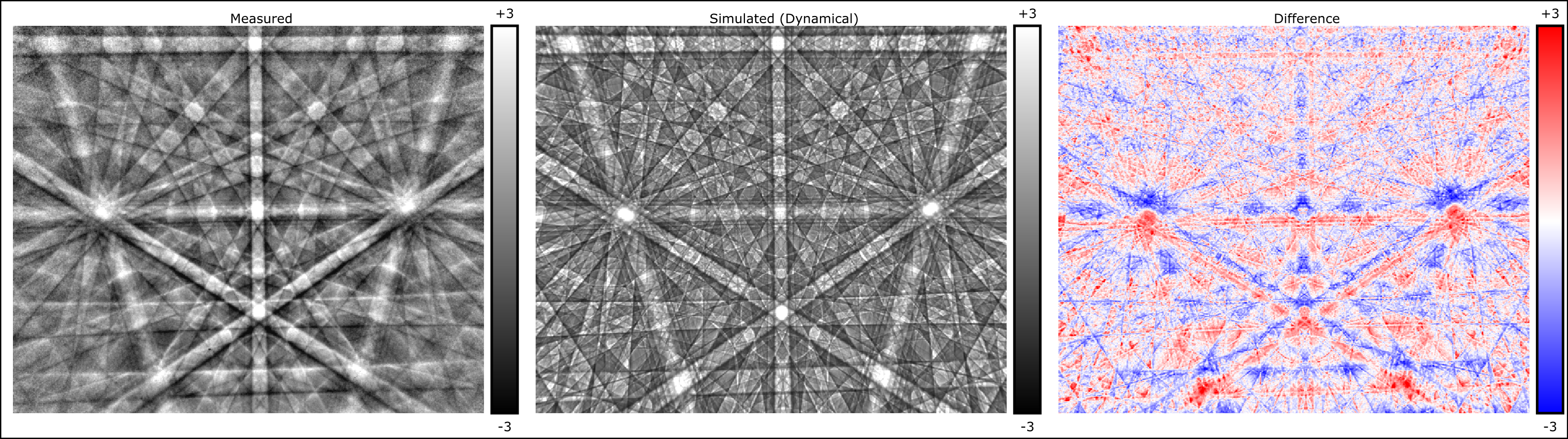}
         \caption{622x512 3x3 R=0.6583}
     \end{subfigure}
     \begin{subfigure}[b]{0.7\textwidth}
        \centering
    \includegraphics[width=\textwidth]{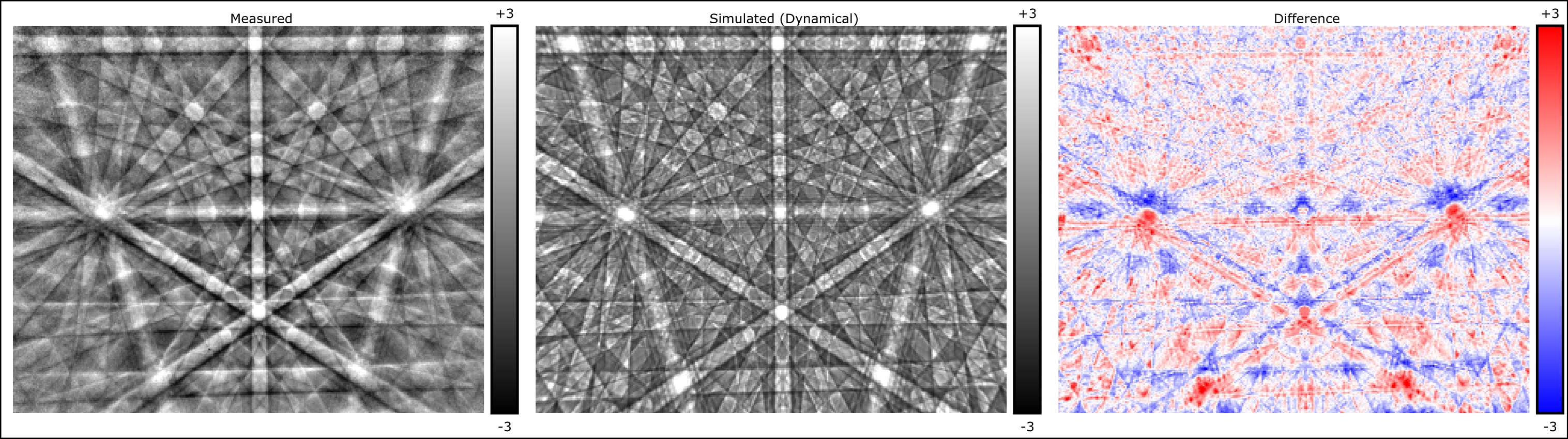}
        \caption{$311\times256$, $7\times7$, R=0.7284}
     \end{subfigure}
     \begin{subfigure}[b]{0.7\textwidth}
         \centering
   \includegraphics[width=\textwidth]{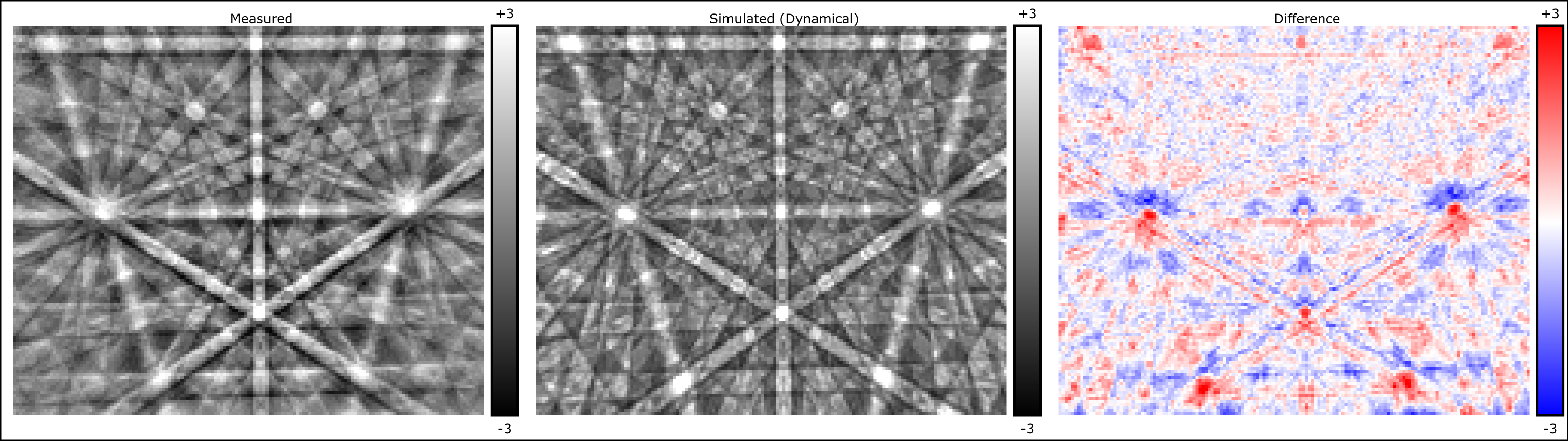}
         \caption{155x128 13x13 R=0.7934}
     \end{subfigure}
    \begin{subfigure}[b]{0.7\textwidth}
        \centering
   \includegraphics[width=\textwidth]{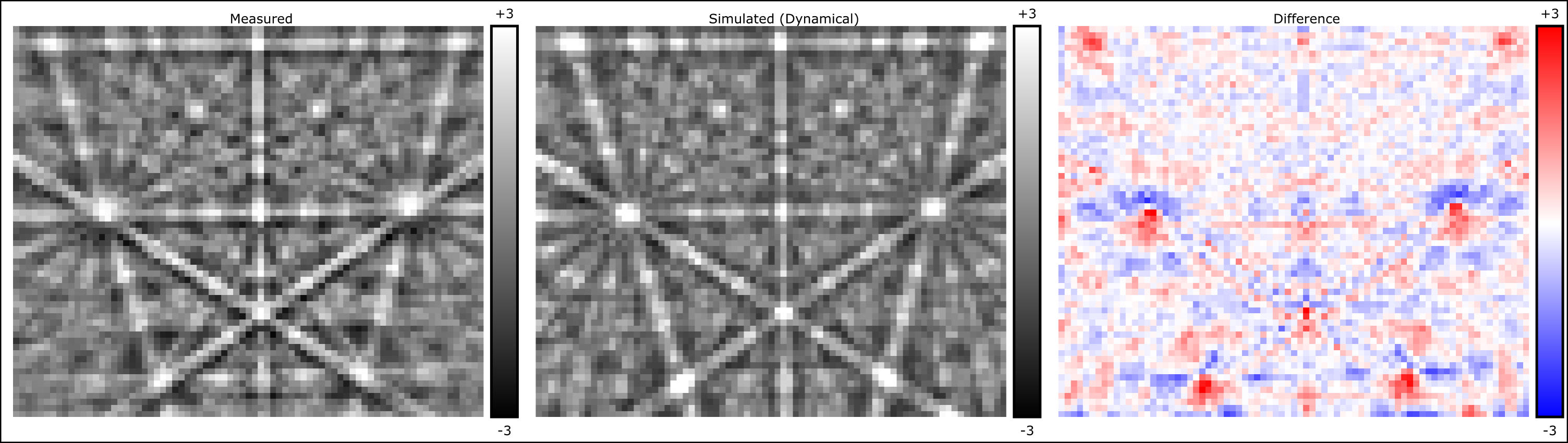}
        \caption{77x64 25x25 R=0.8283}
    \end{subfigure}
    \begin{subfigure}[b]{0.7\textwidth}
        \centering
    \includegraphics[width=\textwidth]{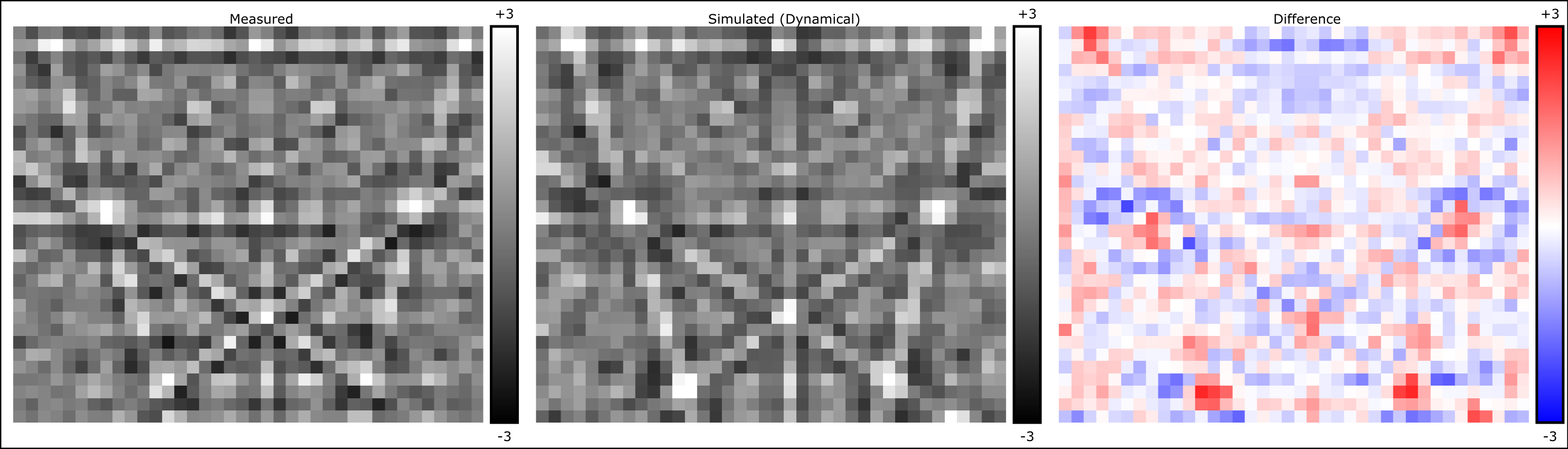}
         \caption{38x32 49x49 R=0.8011}
    \end{subfigure}
    \vspace{-3ex}
    \end{center}
        \caption{Comparison of the effect of binning to different pattern sizes and the resulting correlation coeffcients $R$. The supersampling is set to keep the total number of projected samples approximately constant.}
        \label{fig:patterns}
\end{figure}

In Figure \ref{fig:patterns} we give a visual summary about the loss of pattern resolution that is related to an increasing binning.
We see that the cross-correlation coefficient is initially increasing when decreasing the resolution from Figure \ref{fig:patterns}(a) to (c), which we attribute to the reduction in noise per pixel, while the actual resolution is still oversampling the available exprimental pixel information (i.e. the pixel spacing is well below the feature size in the Kikuchi pattern).
For the resolution of $311\times256$ we can hardly observe any pixelization, which is in line with the numerical results in Table \ref{tab:resu}.
When the resolution reaches $38\times32$, the correlation coefficient is again decreasing, which we could take as an indication that we reach a lower limit of Kikuchi pattern pixel information somewhere in this region of pixel resolution.

An additional type of information results from observation of the appearance of color noise in the $K_3$ maps shown in Figures \ref{fig:si512} and \ref{fig:si128}.
Observing which part of the $K_3$ distribution stays constant, we estimate that a reliable deformation-type characterization requires strain magnitudes larger than about $||\bm{\varepsilon}||_{K_3}^{\mathrm{min}} \approx 1$mm/m due to the inherent precision limits of the experimental strain tensor components. Taking into account that the simultaneous noise level in the strain magnitude is of the order of 0.1 mm/m, we see that the information about the type of deformation already becomes unreliable for strains that are about one order of magnitude above the noise limit.

\FloatBarrier
\subsection{FEM simulations}

\begin{figure}[htb!]
\begin{center}
    \includegraphics[width=\textwidth]{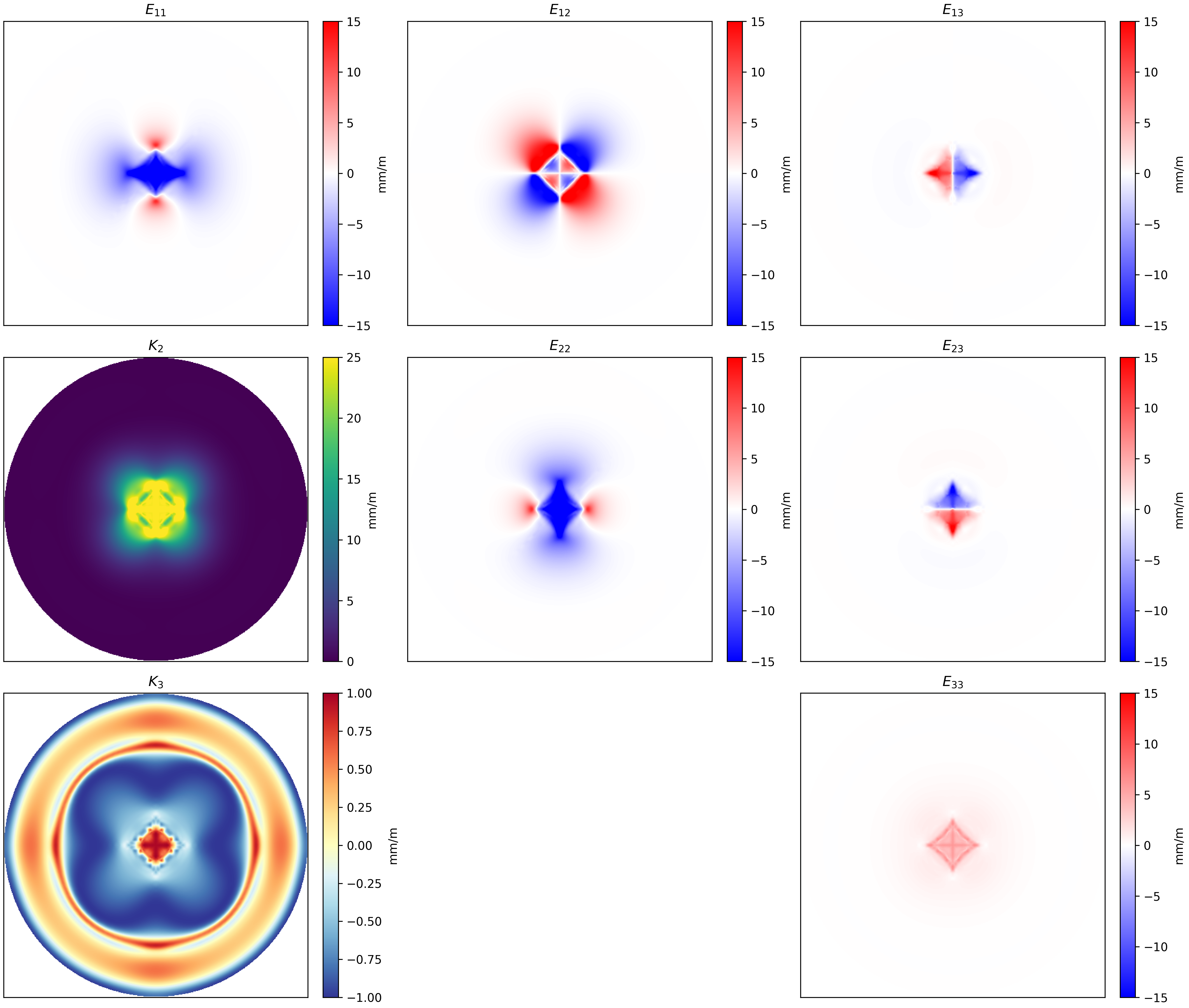}
    \vspace{-3ex}
    \end{center}
        \caption{FEM simulations: components of the strain deviator $E$ and the invariants $K_2$ and $K_3$.}
        \label{fig:fem_deviator}
\end{figure}

Figure\,\ref{fig:fem_deviator} presents results of the FEM simulation that models similar conditions as in the experimental indentation test.
Beyond the immediate central area of the indenter, the components $E_{11}$, $E_{22}$, and $E_{12}$ appear qualitatively accurate and also exhibit the correct magnitude and sign. 
We exclude the central zones close to the indenter tip from the discussion since reliable experimental data cannot be obtained for these areas.
The most significant discrepancy occurs in $E_{33}$: possibly due to topography causing phantom strain in the experiment from the non-planar surface of the indenter crater. The Z-related shears $E_{13}$ and $E_{23}$ being near zero align with the experimental observations of these comparatively low-magnitude components.
$K_2$ qualitatively appears correct and appropriately sized, though details near the crack differ, attributed to strain release by cracking. The trend for $K_3$ towards uniaxial compression in the center is observed experimentally, albeit with more significant deviations towards pure shear as compared to the experiment. The differences observed between the simulated and experimental $K_3$ could stem from phantom strains in $E_{33}$, as well as edge effects in the finite-size simulation model. It should be noted that the developed numerical model replicates the ideal nanoindentation loading setup, including perfect tip geometry and a flawless flat sample surface without any slope or anisotropy in mechanical properties. At the same time, the fracture mechanisms responsible for local relaxations are not considered. Therefore, the numerical model may slightly overestimate the predicted values with respect to the experimental setup with all its uncertainties.The comprehensive assessment of the FEM simulations and their effects on the accuracy of experimental EBSD-derived strain measurements will be explored in future studies.

\section{Summary and Conclusions}

We presented an overview of an EBSD pattern analysis technique that utilizes high-resolution simulations of Kikuchi patterns to evaluate isochoric relative deformation gradient tensors from experimentally obtained Kikuchi patterns of lower resolution. 
As an application example, we performed high-resolution orientation and strain analyses for hardness test indentations on Si(100) wafers, employing Kikuchi patterns of varying resolutions. 

Using simulation-based supersampling, this technique demonstrates noise levels around \mbox{$1 \times 10^{-4}$} for the relative deviatoric strain norm and relative rotation angles in nominally strain-free regions of the silicon wafer. 
A comparison with other published EBSD studies using the approach described in this paper indicates a similar high level of agreement \cite{osborn2018um,ernould2022mc}. 

A limitation of the present approach is its relative nature, necessitating the use of strain-free reference points.
In deformed polycrystalline samples, the approach thus is mainly sensitive only to the small-scale intergranular strains (type III  \cite{withers2001mst}), if strain-free reference points cannot be established in the grains of the polycrystalline material.

A particular finding of the present study is that the precision of the strain tensor components affects the reliability of the information regarding the type of shape change, such as differentiating uniaxial expansion from contraction. 
We find that the determination of the type of shape change already becomes unreliable at strain magnitudes one magnitude higher than the noise limit, i.e. at about $1 \times 10 ^{-3}$ mm/m when the simultaneous noise level in the strain magnitude is of the order of $1 \times 10 ^{-4}$ mm/m as in the present study. 
This precision effect will be generally important in the wider context of EBSD-based strain determination by different methods.

Affirming the fundamental conclusions of previous studies, the method described here offers the potential to significantly reduce the required data size for EBSD pattern analysis by approximately two orders of magnitude. 
This can alleviate the need for super-high-resolution pattern acquisition that has been successfully demonstrated previously using 2k$\times$2k resolution detectors  \cite{wang2021um,ventura2024um,dingley2018icotom,nordif2023hr4m}.
However, such high detector resolutions go hand in hand with excessively large data sets, which can be difficult to handle practically. 

Challenging the common belief that "more pixels are better",  our findings suggest that $256 \times 256$ pixel detectors can probably meet the needs of most EBSD analysis tasks when using pattern matching strategies.

\FloatBarrier
\section*{Acknowledgements}
This work was supported by the Polish National Science Centre (NCN), grant number \\ \mbox{2020/37/B/ST5/03669}.
The research results presented in this paper have been developed with the use of equipment financed from the funds of the "Excellence Initiative - Research University" program at the AGH University of Krakow. We gratefully acknowledge Polish high-performance computing infrastructure PLGrid (HPC Center: ACK Cyfronet AGH) for providing computer facilities and support within computational grant no. PLG/2024/017298.
The dataset with an EBSD pattern resolution of $622 \times 512$ pixels is available at \url{https://doi.org/10.5281/zenodo.14059949}.

\FloatBarrier
\clearpage

\end{document}